\documentclass{ws-ijgmmp}

\usepackage{graphicx}
\usepackage{epstopdf}
\usepackage{amsmath}
\usepackage{amsfonts}
\usepackage{amssymb}
\usepackage{latexsym}
\usepackage{hyperref}
\usepackage[english]{babel}
\usepackage[utf8]{inputenc}
\usepackage{xcolor}
\usepackage{slashed}

\newcommand{\bee}{\begin{equation}}
\newcommand{\eee}{\end{equation}}
\newcommand{\eaa}{\end{eqnarray}}
\newcommand{\baa}{\begin{eqnarray}}

\begin{document}

\markboth{M.J. Neves, E.M.C. Abreu, J.B. OLiveira, M. Kesseles Gon\c{c}alves}
{Thermostatistical analysis for short-range interaction Potentials}

%
\catchline{}{}{}{}{}
%

\title{Thermostatistical analysis for short-range interaction Potentials}

\author{M. J. Neves\footnote{E-mail: mneves@ufrrj.br}}

\address{Department of Physics and Astronomy, University of Alabama, Tuscaloosa, Alabama 35487, USA;\\
Departamento de F\'{i}sica, Universidade Federal Rural do Rio de Janeiro, BR 465-07, 23890-971, Serop\'edica, RJ, Brazil}

\author{Everton M. C. Abreu\footnote{E-mail: evertonabreu@ufrrj.br}}

\address{Departamento de F\'{i}sica, Universidade Federal Rural do Rio de Janeiro, BR 465-07, 23890-971, Serop\'edica, RJ, Brazil;\\
Departamento de F\'{i}sica, Universidade Federal de Juiz de Fora, 36036-330, Juiz de Fora, MG, Brazil and\\
Programa de P\'os-Gradua\c{c}\~ao Interdisciplinar em F\'isica Aplicada, Instituto de F\'{i}sica, Universidade Federal do Rio de Janeiro, 21941-972, Rio de Janeiro, RJ, Brazil}

\author{Jorge B. de Oliveira\footnote{E-mail: jorgebernardo998@gmail.com}}

\address{Departamento de F\'{i}sica, Universidade Federal Rural do Rio de Janeiro, BR 465-07, 23890-971, Serop\'edica, RJ, Brazil}

\author{Marcelo Kesseles Gon\c{c}alves \footnote{E-mail: marcelokesseles@ufrrj.br}}

\address{CAPES-PIBID/F\'{i}sica, Departamento de F\'{i}sica, Universidade Federal Rural do Rio de Janeiro, BR 465-07, 23890-971, Serop\'edica, RJ, Brazil}

\maketitle

\begin{history}
\received{(09 August 2020)}
\revised{(03 September 2020)}
\end{history}

\begin{abstract}
In this paper, we study the thermodynamics of short-range central potentials, namely, the Lee-Wick potential, and the Plasma potential.
In the first part of the paper we obtain the numerical solution for the orbits equation for these potentials. Posteriorly, we introduce the
thermodynamics through the microcanonical and canonical ensembles formalism defined on the phase space of the system.
We calculate the density of states associated with the Lee-Wick and the Plasma potentials. From density of states, we obtain the thermodynamical physical quantities like entropy and temperature as functions of the energy. We also use the Boltzmann-Gibbs formalism to obtain the partition functions, the mean energy and the thermal capacity for these short-range potentials.
\end{abstract}


\section{Introduction}

The Yukawa Potential (YP) in atomic physics, namely, the Debye-H\"uckel potential in plasma physics, is also known as the screened Coulomb potential and it deserves the interest of researches of many areas of physics \cite{yukawa1,yukawa2}. It has its origins in the intention of describing, in a much more precise way, the interactions of forces that act at very small distances.

Originally, we can describe the interactions, for example, of the strong force inside atomic nucleus, i.e., the strong nucleons-nucleons interactions, thanks to the meson exchange, depicted in nuclear physics. Concerning the Coulomb potential having the $K/r$ formulation, where $K$ is any constant, it is good to describe the general force effects, where the limit of the interaction has an almost infinite limit.   We know that if any charge is a very small one, we have a bigger potential and, if the distance is big, we have a small potential. However, the same is not true to evaluate the strong force, since it is measured only at very small distances, which has a very small interaction limit.  To work with these limitations is why H. Yukawa investigated and developed the potential that, today, has his name \cite{yukawa1,yukawa2}.

As we said just above, the YP can also be used to show a screened Coulomb potential due to the cloud of electronic charges around the nucleus in atomic physics. Or to represent the shielding by outer charges of the Coulomb field that an atomic electron feels in hydrogen plasma. Although the Schr\"odinger equation concerning this potential cannot be solved exactly, there are various numerical and perturbation methods that have been thought up in order to capture the energy levels and related physical quantities \cite{ba}.

It is very clear to theoretical physicists that there is a connection between the field equations of gravity and the first law of thermodynamics on the boundary of spacetime \cite{jacobson,padmanabhan,egj}.   This correspondence leads us to the entanglement between thermodynamics and gravitation, which has been explored since the work of Jacobson \cite{jacobson}.   Recently, the proposal of Verlinde \cite{verlinde} conjectures that gravity is an emergent phenomenon due to the increasing of entropy.   This emergence occurs between holographic screens \cite{cco}, which is a fundamental ingredient together with entropy.   Its importance can be seen through several corrections to the entropic-area expression, which introduces several modifications into gravity theories and their respective cosmology
\cite{entropic-cosmol,unruh-temp}.   Since it was demonstrated \cite{unruh-temp} that, if the Unruh temperature \cite{unruh} is connected to the holographic screen, so the quantum statistics of the surface degrees of freedom can help us to obtain a theoretical cornerstone concerning the well known Modified Newtonian Dynamics (MOND) theory \cite{milgrom}.

To explain the accelerated expansion of the Universe, thermodynamical pictures have been described.   One of them is the Bekenstein-Hawking entropy, where this last one is proportional to the surface area of the event horizon \cite{bekenstein,hawking}.   The other important scenario is the holographic principle (HP), in which the information of the bulk is stored on the horizon \cite{princ-holog}.  The HP was used to propose that gravity is an entropic force obtained from changes in Bekenstein-Hawking entropy \cite{jacobson,verlinde,padmanabhan}.   To demonstrate this concept, the cosmological equations have been widely investigated in an homogeneous and isotropic Universe \cite{cosmol-equat}.    However, the cosmological constant was not discussed.   Alternatively, in \cite{efs} the author suggested an entropic cosmology that takes into account the neglected surface terms on the horizon of the Universe \cite{varios,bps,bs,komatsu-varios,dg,dgp}.   In this so-called entropic cosmology, the explanation of the accelerated expansion was obtained from the entropic forces on the horizon of the Universe.

Thermodynamical analysis were carried out for different potential in the last few years.   In \cite{29}, the authors analyzed the thermodynamics of the Schroedinger equation with a Deng-Fan-type potential.   In \cite{30}, it was studied the Klein-Gordon approach or particles embedded in exponential-type molecule potential together with their thermodynamical characteristics in $D$ dimensions.   The harmonic oscillator plus an inverse square potential were investigated thermodynamically in \cite{31}.   Concerning different diatomic molecules, their thermodynamical properties were analyzed in \cite{32,33} and with general molecular potential in \cite{34}.   In \cite{oiioc}, the authors solved the Schroedinger equation using the modified factorization method for the modified Yukawa potential, also known as generalized inversely quadratic Yukawa potential.   Hence, we can see the importance of the thermodynamics analysis in any kind of interaction.

In this work we have analyzed the thermodynamical aspects of several different short-range potentials,  The microcanonical and canonical formalisms were carried out and compared.   Thermodynamical and statistical mechanics  quantities like entropy, temperature were computed together with the partition function, mean energy and heat capacity, respectively.


We have organized the discussion of these issues in such a way that in section 2 where we have described some classical aspects of short-range potentials such as force and equations of motion.   In section 3 we analyzed the thermodynamical aspects of these short-range potentials and in section 4, their statistical mechanics results.   In section 5, we made some discussions, conclusions and perspectives.


\section{The orbit equation for short-range forces}
Differently from the Coulomb potential, the Lee-Wick (LW) \cite{LeeWick69,LeeWick70,Connell08} short-range potential shows an association with the involved particles' masses. In other words, LW realized that the interactions, in general, are inversely proportional to the potential mediator's mass. In this way, he has developed a potential that has the following form\footnote{We are considering the natural units $\hbar=c=1$.}:
\begin{eqnarray}\label{LW}
V_{LW}(r) = K \, \frac{1-e^{- \mu \, r}}{r} \; ,
\end{eqnarray}
\noindent where $K$ is the potential strength and $\mu=Mm/(M+m)$ represents the reduced mass between the two particles masses, $M$ and $m$.
Note that the LW potential is the subtraction of the Coulomb with the Yukawa potential. Furthermore, when $\mu r \ll 1$, the potential
function is finite at the origin, and its value is $V_{LW}(r=0)=K\mu$.

We can see clearly that the potential is monotone increasing in $r$ and after a certain point, it turns to be negative, which means that the force goes to be attractive. It is worth mentioning that, relative to the interactions between both a meson field and a fermion field, the constant $K$ has the same value as the gauge coupling constant between both fields.

Concerning the Coulomb potential case, the reduced mass goes to zero, which represents a zero mass boson as being the potential mediator, and $K$ can have positive or negative real values. Specifically, for $K > 0$ it has been found in the literature that the S-wave bound
states exist only for values of $\mu$ below just a determined critical value
$\mu_c \sim 1.1906\, A$ in atomic units \cite{go}. Moreover, differently from the Coulomb potential, the Yukawa one permits the appearance of resonant states.
Questions about the solution of the Schr\"odinger equation for this potential were investigated intensely in the past by using several numerical and perturbative analyzes.
The reason is that it is not possible to find exact analytical solutions since there is a lack of shape invariance relative to this potential \cite{varios-0}.
Considering the short-range behavior of the YP due to the decaying exponential term (i.e., $e^{- \mu r}$), there is a singularity at the origin thanks to $r^{-1}$.  The centrifugal term $r^{-2}$ behavior, together with the exponential term, both are responsible for the non-trivial task of
obtaining accurate numerical solutions.   To provide a thermodynamical investigation, in this work we will analyze the short-range interaction problem using, among others, the YP.

\subsection{The central potential: a quick review}

Since a central force $F(r)$ is conservative, it can be written as minus the radial derivative of a $U(r)$ potential function.
%
%
The motion of two particles of masses $M$ and $m$, that interact under the action of central force $F(r)$, is described
by the radial equation
\begin{eqnarray}\label{Eqmotion}
\mu \, \ddot{r} = -\frac{dU_{eff}}{dr} \; ,
\end{eqnarray}
where $r$ is the relative coordinate between $M$ and $m$, $U_{eff}(r)$ is the effective potential that can be written as
\begin{eqnarray}
U_{eff}(r) = U(r) + \frac{L^{2}}{2 \mu r^{2}} \; ,
\end{eqnarray}
where the quantity $L^2$ means the conserved angular momentum of a general two-particle system. Since we know the form of the potential function
$U(r)$, the equation of motion in Eq. (\ref{Eqmotion}) can be integrated to obtain the radial solution as a function of time.
In the case of an interaction force system, it is well known that the trajectory carried out by $m$ interacting with $M$ is given by
\begin{eqnarray}
\frac{d^{2}r}{d\theta^{2}} - \frac{2}{r}\left(\frac{dr}{d\theta}\right)^{2} = r + \left(\frac{\mu r^{4}}{L^{2}}\right) F(r) \; .
\end{eqnarray}
To solve the previous equation in order to ease the calculations, we will provide a variable transformation of the kind
$u= 1/r$. In this way, the final form of the equations of motion is
\begin{eqnarray}
\frac{d^{2}u}{d\theta^{2}} + u = -\frac{\mu}{L^{2} u^{2}} \, F\left(\frac{1}{u}\right) \; .
\end{eqnarray}
Therefore, if we know the form of both functions $U(r)$ and $F(r)$, we can determine the information about the behavior of $m,\,M$ and
its corresponding trajectory. We will use several potentials in the next subsections.
%


\subsection{The orbit solution for the Lee-Wick potential}

\begin{figure}[t]
 \centering
 \includegraphics[width=0.49\textwidth]{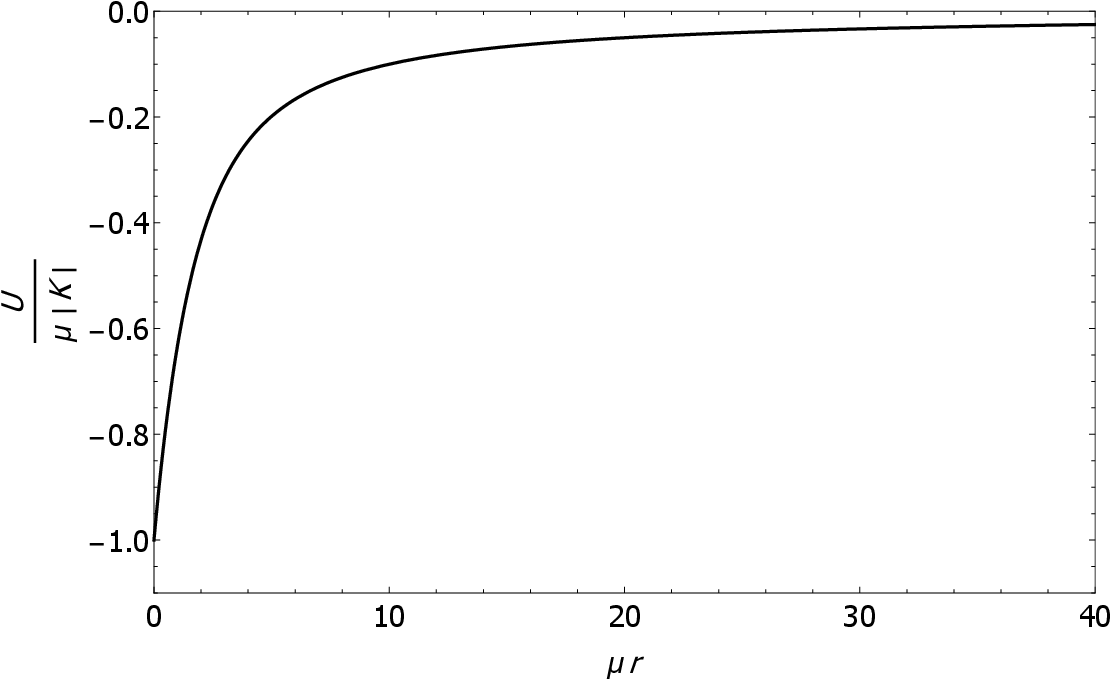} 
 \includegraphics[width=0.49\textwidth]{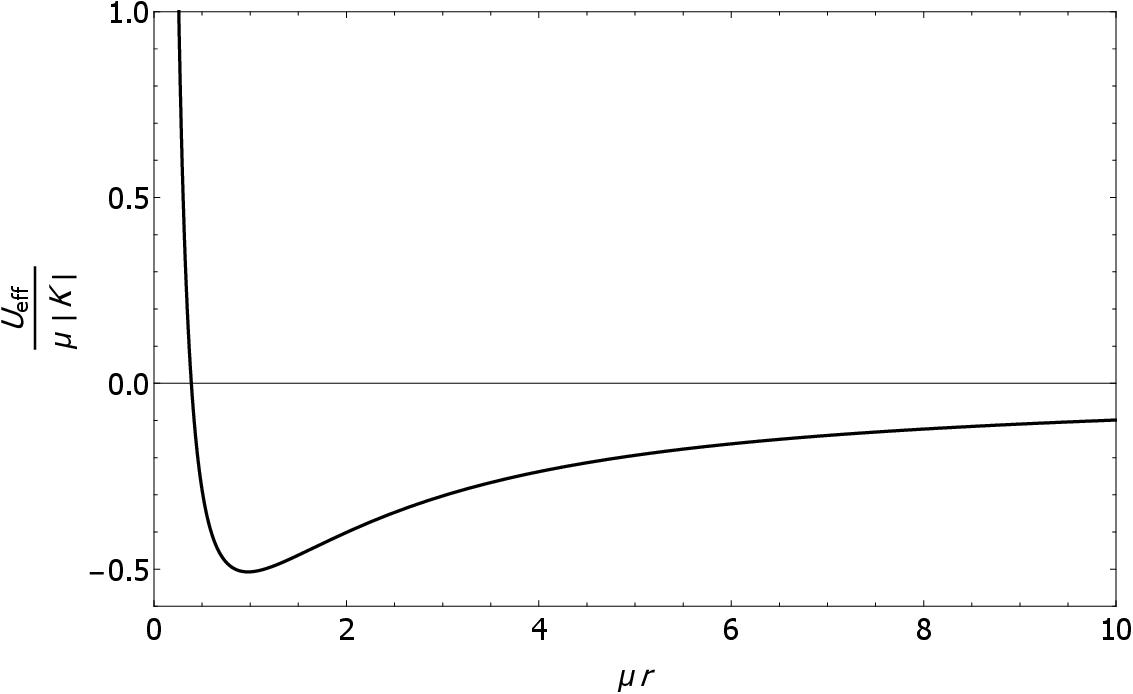}
\caption{
Left panel: Plot of Lee-Wick potential by $\mu |K|$ units versus the dimensionless $\mu r$.
Right panel: The Lee-Wick effective potential as function of the radial distance.} \label{UplotsLW}
\end{figure}

To analyze the kind of interaction that comes from the Lee-Wick (LW) potential, we can say that the force generated by it is, considering the work and energy theorem, a central force.  We can say the same concerning all the potentials that will be considered here.
Having said that, we can use the relation between force and potential energy. Considering the LW case in Eq. \eqref{LW},
%
%
the corresponding effective potential is given by
\begin{eqnarray}
V_{eff}^{(LW)}(r) = -|K| \, \frac{1-e^{- \mu \, r}}{r} + \frac{L^{2}}{2 \mu r^{2}} \,\, ,
\end{eqnarray}
and its behavior is shown in figure (\ref{UplotsLW}).
If we analyze the effective potential energy of the system, under the same conditions used before, we could obtain the points of maximum distance and maximum proximity, since we consider that at these points it does not exist kinetic energy. Hence, from
\begin{eqnarray}
V_{eff}^{(LW)}(r) \simeq \frac{L^{2}}{2 \mu r^{2}} + \frac{K}{r}-\mu \, K \; ,
\end{eqnarray}
where these distances are obtained by making $V_{eff}^{LW}(r)=-|E|$, where we can consider as negative the total energy of system.
If we make the necessary calculations, we have that
\begin{eqnarray}
\frac{1}{r_{\pm}} = \frac{\mu \, |K| }{L^{2}} \, \pm \sqrt{\frac{\mu^{2} |K|^{2}}{L^{4}} - \frac{2\mu}{L^{2}}\left(|E| + \mu |K|\right)} \,\,,
\end{eqnarray}
and the turning points are
\begin{eqnarray}
\frac{1}{r_{p}} &=& \frac{\mu \, |K|}{L^{2}}\left(1\,+\,\sqrt{1\,-\,\frac{2\,L^{2}}{\mu \,|K|^{2}}(|E| + \mu |K|)}\,\right)\;,
\\
\mbox{} \nonumber \\
\mbox{} \nonumber \\
\frac{1}{r_{a}} &=& \frac{\mu \, |K|}{L^{2}}\left(1\,-\,\sqrt{1\,-\,\frac{2\,L^{2}}{\mu \,|K|^{2}}(|E| + \mu |K|)}\right) \; ,
\end{eqnarray}
%
%
%
%
%
where $r_{p}$ is the point of maximum approximation, and $r_{a}$ is the point of maximum distance.
Namely, considering the point $r_{p}$, we have the smaller distance between $M$ and $m$ and
and the point $r_{a}$ is just the opposite, where they will be at a maximum distance, i.e., they still interact,
and after that point, they will not.

The radial interaction force generated by the LW potential is given by
\begin{eqnarray}
F_{LW}(r) = -\frac{K}{r^2}+\frac{K}{r^2} \, e^{- \mu r} \, \left( \,  1 + \mu \, r \, \right) \; ,
\end{eqnarray}
%

%
%

\noindent where, if we consider $K>0$, we have that $F(r)>0$, which characterizes a repulsive force.  On the other hand, if the $K$ value is negative, the value of $F_{LW}(r)$ will be negative, which characterizes an attractive force.

As we said before, the Yukawa potential was created to describe the very small distance (short-range) interactions, i. e., $\mu r \ll 1$.  Hence, considering the limit of the interaction force as being
\begin{eqnarray}
F_{LW}(r) \simeq - \frac{|K| }{r^{2}} + \frac{|K| }{r^{2}} \left( 1\,-\,\mu^2 r ^2 \right)\simeq-\mu^2 |K| \; ,
\end{eqnarray}

\noindent and by solving this limit, we obtain that, for distances that have a tendency of being very small.   The interaction force tends to have a very strong interaction such that this type of force, when $r \rightarrow 0$, increases as the bodies are closer.  Moreover, if $r \rightarrow \infty$ we can verify that the interaction force goes to zero, since it has a negative exponential function.


To use the interaction force of the LW potential we have, firstly, to change the same variable $r$ and we will have a new variable $u$ such that
$u\,=\, 1/r$. Hence, the orbit equation will be constructed as
%
\begin{eqnarray}
\frac{d^{2}u}{d\theta^{2}} + u = -\frac{\mu K}{L^{2}} \, \left[ -1+ \left(1+\frac{\mu}{u} \right)e^{-\frac{\mu}{u}} \right] \; .
\end{eqnarray}
\begin{figure}
\centering
\includegraphics[scale=0.45]{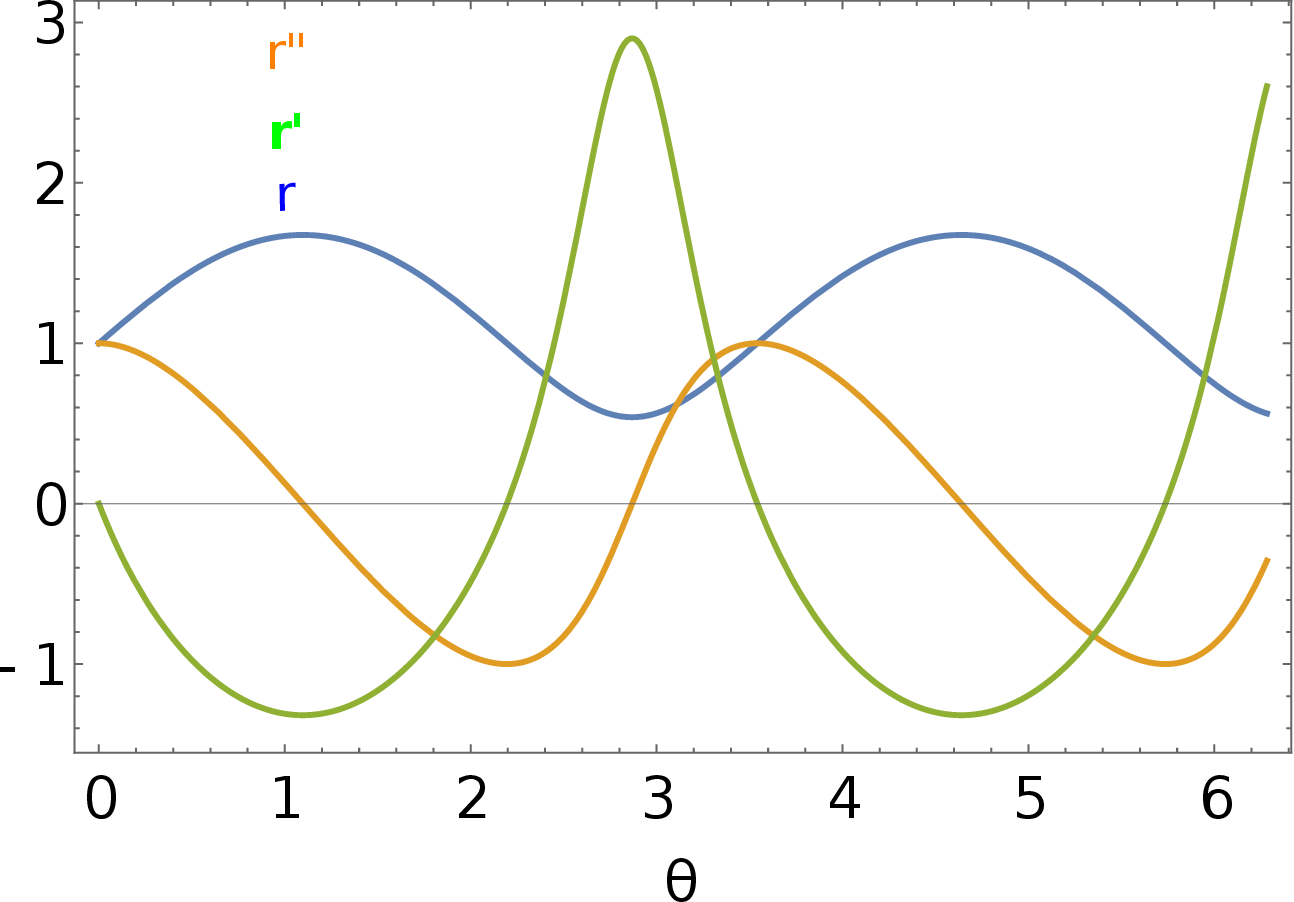}
\caption{The radial coordinate $r$, it first and second derivative in relation to $\theta$, as function of $\theta$.}\label{SolDiffLW}
\end{figure}
\noindent Since the LW potential works only in the nuclear scale,
$r\rightarrow 0$, we will adopt an approximation such that
$\mu r \ll 1$ and the reduced mass, at the nuclear level is very small, just like the distance between them.
%
In this case, the orbit equation turns out to be a non-linear differential equation given by
\begin{eqnarray}
\frac{d^{2}u}{d\theta^{2}} + u \simeq \frac{\mu^3 \, K}{L^{2} \, u^2} \; .
\end{eqnarray}


For this second order differential equation, the analytical solution has the form
\begin{equation}
u(\theta) \, = -\frac{\mu \, K}{L^{2}} + A \, \cos\left( \theta-\theta_{0}\right)
-\frac{\mu^{3} \, K}{L^{2}}\left[ \, \cos(\theta) \, \mbox{cosintegral}(\theta)  + \mbox{sen}(\theta) \, \mbox{senintegral}(\theta) \, \right] \; ,
\end{equation}
%
and this solution, using a ($r-\theta$) plane, is shown (blue line) in figure (\ref{SolDiffLW}). In the same figure, the first derivative $r'$ (green line) and the second one $r^{\prime \, \prime}$ (orange line) are also shown as a function of the $\theta$-angle.


\subsection{The orbit solution for Debye-H\"uckel potential}

The Debye-H\"uckel potential, known in plasma physics, can be represented by the central potential
\baa \label{plasmaU}
V_{DH}(r)=-V_{o} \, \Big( 1 \, - \, \frac{a}{r} \, e^{-\alpha r} \Big)^2 = - \, \frac{A}{r^2} \, e^{-2\alpha r} \, + \, \frac Br \,e^{-\alpha r}\,-\,C \; ,
\eaa
where $a$ is a constant carrying the length dimension, $A=a^2\,V_o$, $B=2 \, a \, V_{o}$ and $C=V_{o}$, $V_{o}$ is the coupling strength and $\alpha$ is the potential screening parameter with length dimension. The effective potential associated with the one in Eq. (\ref{plasmaU}) is shown in figure (\ref{Ueffplasma}). The value of the parameters $A,\,B$ and $C$ can represent several potentials. For example, if $A=C=0$ and $B\,\rightarrow\,-B/2$, we have the standard Yukawa potential. For $\alpha=0$ we have the Mie-type potential. For $B=C=0$ and $A=-V_o$, the inversely quadratic Yukawa potential.  And for $A=C=0$ and $\alpha=0$, the Coulomb potential.

The Mie potential was suggested by Gustav Mie in 1903 \cite{mie}. It deals with intermolecular pair potential between two particles at a distance $r$.   It is a very used approximation concerning $n$-body interactions.

The modified Yukawa potential is asymptotic to a finite value as $r \rightarrow 0$ and turns out to be infinite at $r=0$ \cite{oo}.   Its application in many areas of physics was investigated recently \cite{hgzh}. The Yukawa potential can be used in plasma physics, {\it i.e.}, the so-called Debye-H\"uckel potential. In solid state physics and atomic physics, it is known as the Thomas-Fermi or screened Coulomb potential. In nuclear physics, it is the dominant central part of nucleon-nucleon interaction.
\begin{figure}[t]
\centering
\includegraphics[scale=0.35]{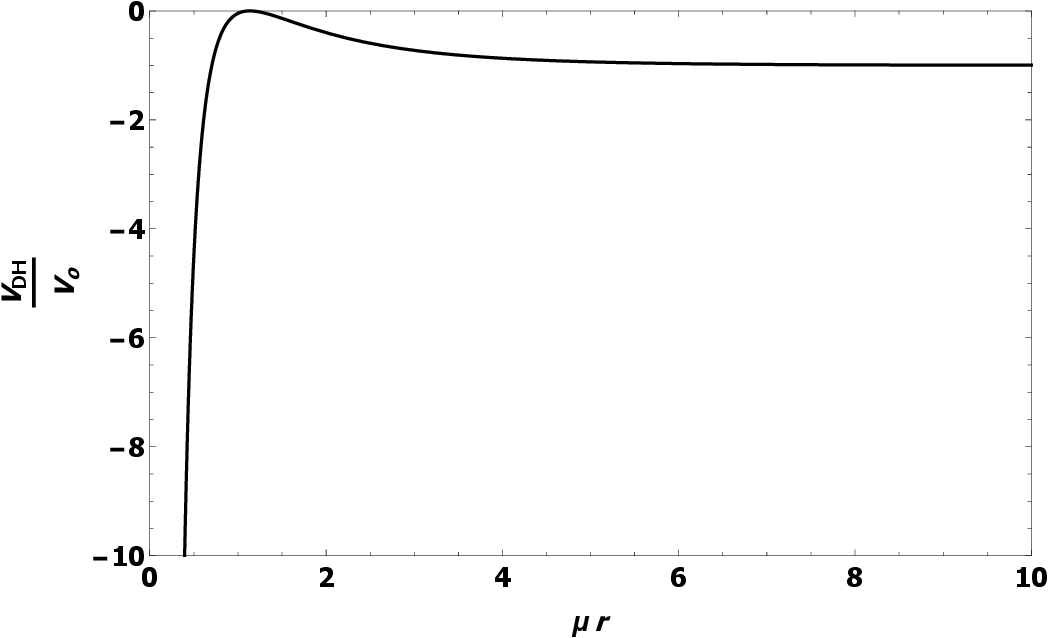}
\includegraphics[scale=0.35]{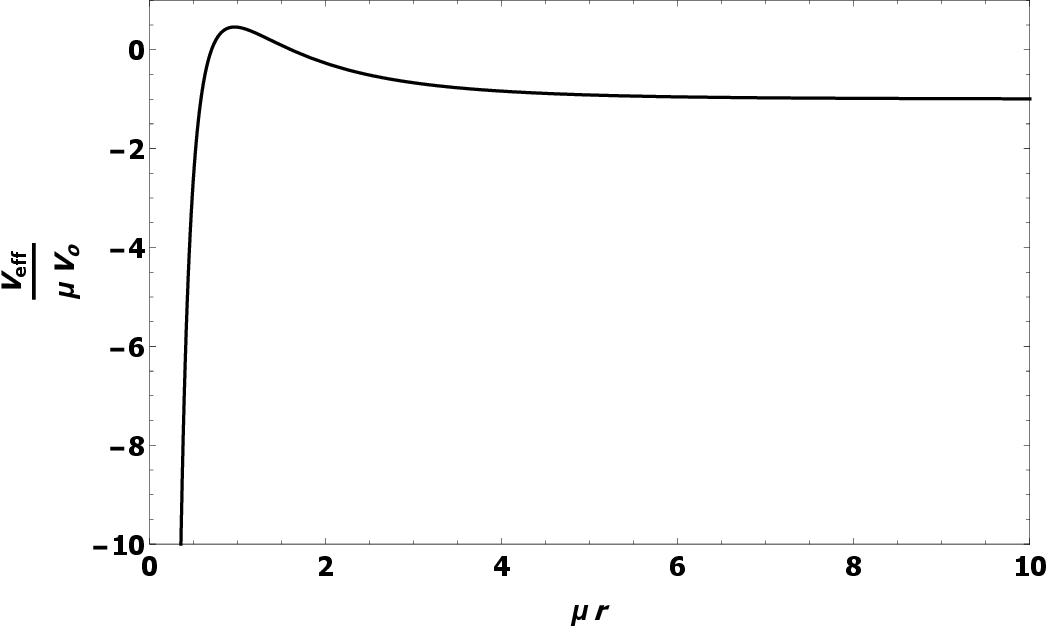}
\caption{Left panel : The DH potential by unity of $V_{0}$ as function of dimensionless $\alpha r$.
Right panel : The effective DH potential by unity of $V_{0}$ as function of dimensionless $\alpha r$.}\label{Ueffplasma}
\end{figure}
The central force associated with the potential in Eq. (\ref{plasmaU}) is given by
\begin{eqnarray}\label{FrPlasma}
F_{DH}(r)=-\frac{2 \,a\, V_{o}}{r^2} \, e^{-\alpha r} \left(1+\alpha r\right) \left(1+\frac{a^2\,e^{-\alpha r}}{r}\right) \; ,
\end{eqnarray}
where the equation of motion for this central force can be written as
\begin{eqnarray}\label{EqPlasma}
\frac{d^{2}u}{d\theta^{2}} + u = \frac{2 \, \mu \,a\, V_{o}}{L^{2} \, u} \, e^{-\alpha/u} \left( \, 1+\frac{\alpha}{u} \, \right)
\left( \, 1+a^2\,u\, \, e^{-\alpha/u} \, \right) \; .
\end{eqnarray}
Note that, for large distances, $\alpha r \gg 1$, the force in Eq. (\ref{FrPlasma}) goes to zero.  But for very short distance $\alpha r \ll 1$,
we obtain the approximated result
\begin{eqnarray}
F_{DH}(r) \simeq \frac{2 \, K}{\alpha^2 \, r^3} \; ,
\end{eqnarray}
and the equation of motion in Eq. (\ref{EqPlasma}) can be reduced to standard differential equation
\begin{eqnarray}\label{EqPlasmas}
\frac{d^{2}u}{d\theta^{2}} + a^2 \, u \simeq 0 \; ,
\end{eqnarray}
where $a=\sqrt{1+2 \, \mu \, V_{o}/(\alpha L)^2}$. The solution of Eq. (\ref{EqPlasmas}) is well known and it is given by
\begin{equation}
u(\theta) \simeq A \, \cos\left[ \, a \, \left( \theta-\theta_{0} \right) \, \right] \;  .
\end{equation}

For the Mie potential, the force is given by
\bee
\label{mie}
F(u) \, = \, - \, 2 \, a \, V_{o} \, u^2 \, \Big(1\,+\,a^2\,u \Big) \; ,
\eee
\noindent and the dynamical equation is
\bee
\label{mie-orbit}
\frac{d^{2}u}{d\theta^{2}} + u = \frac{2 a \, \mu \, V_o}{L^2} \, \Big(1\,+\,a^2 u \Big)\,\,.
\eee

Considering the Yukawa potential,
\bee
\label{yuk}
F(u) \, = \, a \, V_{o} \, u^3 \, e^{-\alpha/u} \, \Big(1 \, + \, \frac{\alpha}{u} \Big) \,\,,
\eee
and
\bee
\label{yuk-orbit}
\frac{d^{2}u}{d\theta^{2}} + u = -\,\frac{a V_o u}{L^2} e^{-\alpha/u} \Big(1\,+\,\frac{\alpha}{u} \Big)\,\,.
\eee
\bigskip

Finally, for the inverse quadratic Yukawa potential
\bee
\label{inv-yuk}
F(u) \, = \, - \, u^3 \, e^{-\alpha/u} \Big(1 \, + \,\frac{\alpha}{u} \Big)\,\,,
\eee
and
\bee
\label{inv-yuk-orbit}
\frac{d^{2}u}{d\theta^{2}} + u = \frac{\mu u}{L^2} \, e^{-\alpha/u} \, \Big(1\,+\,\frac{\alpha}{u} \Big)\,\,.
\eee

After these classical results we could ask what would be their behaviors if we intend to introduce quantum features into these potentials. Or a semi-classical analysis would be more physically profitable.  One way to introduce this semi-classical features is through the analysis of the behavior of these potentials in noncommutative phase-spaces.  It is an ongoing research which will published elsewhere.

From now on we will investigate the thermostatistical features of the potential discussed above.  We will use both the microcanonical and canonical ensemble formalisms.


\section{Thermodynamics}

In this section we will deal with the thermodynamics of the potentials described in the last section, which are classified as short-range interaction systems, as we are saying from the beginning.  In the cases analyzed here, additivity implies that the entropy density $S$ is a concave function in $E$.   In short-range systems, it is usual to obtain the microcanonical description from the canonical ensemble equivalence.  The reason is that in that case, one needs to work with some simple integrals over the Boltzmann weights.

In Boltzmann, canonical ensemble, short-range scenarios, the kinetic and thermodynamics temperatures are equal and the specific heat is necessarily positive.  In one dimensional constructions, they should not allow phase transitions if the interaction is short-ranged.   Considering short-range forces, both energy and entropy are additive over the subsystems.  It is well known that there is a lead that ergodicity and mixing apply to the majority of the nonintegrable systems with short-range forces.  To foretell the behavior of structures with short-range forces, we can depend on both  thermodynamics and statistical mechanics, which dissatisfy systems with long-range forces \cite{lprtb}.

If the interaction potential is short-range, each component will interact only with the component that are inside the range of the interaction potential, as we will see.  The short-range forces are restrained to a box to have a nontrivial thermodynamics.  Short-range force systems are extensive.  Considering neutral two particles plasmas, the Debye screening points to an effective short-range interaction potential.  Hence, the equilibrium state of neutral electrolytes and plasmas, for this reason, can be discussed using the standard Boltzmann-Gibbs statistical mechanics.

\subsection{Microcanonical and canonical ensembles}

To construct the statistical description of the systems analyzed here, let us begin with the construction of the microcanonical ensemble depicting the system.   If we have that the Hamiltonian of the system is given by $H(q_i, p_i)$, hence, the volume $g(E)$ of the constant energy surface $E=H(q_i,p_i)$ is an important object in the micro-canonical ensemble.   So, with $g(E)$ we can calculate the entropy and the temperature of the system, as we will show in a moment.

All the relevant thermodynamics characteristics of the system can be fathomed from the $T(E)$ curve.   The microcanonical ensemble exhibits negative heat capacity.   In that scenario, in the case of self-gravitating systems, for example, which is a model of underlying interest, and which is a place where the concepts of statistical mechanics and thermodynamics can be used, the main object of application is astrophysics.   But, as we said before, the general formalism can be utilized in other fields of physics.   Hence, it is very healthy to describe systems in parallel to demonstrate the differences and similarities between a canonical (fixed $T$) and microcanonical (fixed $E$) evolution.   The first difference can be explained by saying that the microcanonical ensemble admits negative specific heats.   On the other hand, the canonical ensemble allows only positive specific heats regions on the caloric curve $T(E)$.   These are called {\it ensembles inequivalence}.   The region of negative heat capacities in the microcanonical ensemble is substituted by an isothermal phase transition in the canonical ensemble.

Since we are considering two-body systems, let us consider two particles depicted by a Hamiltonian of the form
\bee
\label{t-1}
H(\vec{P},\vec{Q},\vec{p},\vec{r})\,=\,\frac{\vec{P}^{\,2}}{2M}\,+\,\frac{\vec{p}^{\;2}}{2\mu}\,+\,V(r) \; ,
\eee

\noindent where $(\vec{Q},\vec{P})$ are, respectively, the coordinates and momenta of the center of mass.   And $(\vec{r},\vec{p})$ are the relative coordinates and momenta of the particle, $M$ is the total mass and $\mu$ is the reduced mass.   As we said above, this system consists of two particles, with different or equal masses, interacting via $V(r)$.   We will presume that the quantity $r$ fluctuates in the interval $(b,R)$.   It is tantamount to suppose that the particles are hard spheres of radius $b/2$.   The system is confined to a spherical box of radius $R$.   We will investigate the ``statistical mechanics" for some toy models, namely, a two-body system confined to different interacting potentials.

Let us start by considering the volume $g(E)$ \cite{padbanabahn4}  associated with the constant energy surface $H=E$ of the phase space.   The density of states is defined as
\bee
\label{t-2}
g(E)\,=\,\frac{1}{N!}\,\int \delta\Big[E-H(\vec{P},\vec{Q},\vec{p},\vec{r})\Big]\,d^3\vec{Q}\, d^3\vec{P}\, d^3\vec{r}\, d^3\vec{p} \; ,
\eee

\noindent which gives the density of states with energy $E$ and $N!\,g(E)$ is the volume of phase space occupied by the constant energy surface itself.  It is related logarithmically to the thermodynamic entropy $S$ of the system, as we will see just below.   This is the term of underlying importance in the microcanonical ensemble.

After integrating over the variables $Q$, $r$ and $P$, we have that
\bee
\label{t-3}
g(E)\,=\,A\,R^3 \int^{r_{max}}_b r^2 dr\,\Big[E\,-\,V(r)\Big]^2\,\, ,
\eee

\noindent where $A=64\pi^5 m^3/3$ and we have considered that both particles have equal masses $m$ and radius equal to $b/2$.

Since $g(E)$ is defined positive, the range of this last integral should be limited to the region where the expression inside the square brackets is also positive.

The entropy $S(E)$ and the temperature $T(E)$ of the system can be defined by the relations
\bee
\label{t-4}
S(E)\,=\,\ln g(E) \quad \qquad
\mbox{and} \quad \qquad
\frac{1}{T(E)}\,=\,\beta(E)\,=\,\frac{\partial S(E)}{\partial E} \; ,
\eee

\noindent and all the relevant thermodynamics properties of the system can be fathomed from the $T(E)$ curve \cite{padbanabahn4} .  The function $T(E)$, being a function of the energy, establish the well known microcanonical caloric curve \cite{chavanis}.    The entropy is defined only up to an additive constant.

All the potentials analyzed here are the ($e^{-\alpha\,r}/r$)-type, which means that, when $r \rightarrow 0$, the exponential term goes to the unity value faster than the $r$-denominator goes to zero.   Hence, this kind of potentials are very similar to the gravitational $r^{-1}$ one.

The investigation of the thermodynamical scenarios of relativistic and nonrelativistic structures has been a useful point of view concerning the comprehension of a great quantity of physical phenomena \cite{amn}.   We can enumerate some examples like nuclear mater, cosmology, QGP in heavy ions collision and etc.  Let us consider some potentials and their respective thermodynamical quantities.  The interaction potential was assumed to be a hard sphere at short distances and decreasing at long distances, like $e^{-\alpha r}$ and $r^{-\alpha}$.


\subsection{The density of states and temperature for Lee-Wick}
%

%

Using the LW potential in Eq. (\ref{LW}) in the integral in Eq. \eqref{t-3}, we obtain the density of states as function of the energy $E$
\baa
\label{LeeWick}
g_{LW}(E)&=& -\,\frac{2AR^3 K^2 E}{\mu} \Bigg\{\frac{\mu K}{6 E^2} \Bigg[ \Bigg(1\,-\,\frac{Er_{max}}{K}\Bigg)^3\,-\,\Bigg(1\,-\,\frac{E b}{K}\Bigg)^3 \Bigg] \nonumber \\
&-&e^{-\mu r_{max}} \Bigg[ \frac 1E \Bigg(1\,-\,\frac{e^{-\mu r_{max}}}{4}\Bigg)\,-\,\frac 1K\,\Bigg(r_{max}\,+\,\frac{1}{\mu} \Bigg) \Bigg] \nonumber \\
&-&e^{-\mu b} \Bigg[ \frac 1E \Bigg(1\,-\,\frac{e^{-\mu b}}{4}\Bigg)\,-\,\frac 1K\,\Bigg(b\,+\,\frac{1}{\mu} \Bigg) \Bigg] \Bigg\}\,\,,
\eaa
\noindent where, from Eq. \eqref{LW} we have that $V_{LW}(r=r_{max})=E$, hence,
\baa
\label{t-6}
r_{max} &=& \frac{\sqrt{2}}{\mu} \sqrt{\mu\,-\,\frac{E}{K}}\,\,, \qquad \mbox{for} \qquad \frac{1-e^{-\mu b}}{b} < \frac{E}{K} < \frac{1-e^{-\mu R}}{R} \nonumber \\
&\mbox{and}& \qquad r_{max}\,=\,R\,\,, \qquad \mbox{for} \qquad \frac{1-e^{-\mu R}}{R} < \frac{E}{K} < \infty\,\,.
\eaa
From Eqs. \eqref{t-4}, the entropy of the LW potential is given by $\ln g(E)$, and the inverse of the temperature is
\baa
\label{beta-lee-wick}
\frac{1}{T_{LW}(E)}
&=& \frac 1E\,-\,\frac{2AR^3K^2 E}{\mu g_{LW}(E)}\Bigg\{-\,\frac 2E A_1(E)\,+\,\frac{1}{\mu K r_{max}}\,A_2(E) \nonumber \\
&-&\frac{\mu}{2E^2}\Bigg[ \Big(r_{max}\,-\,\frac{E}{\mu^2 r_{max}} \Big) \Big( 1\,-\,\frac{Er_{max}}{K}\Big)^2\,-\,b\Big(1\,-\,\frac{Eb}{K}\Big)^2 \Bigg] \nonumber \\
&+&\frac{e^{-\mu r_{max}}}{E^2}\Bigg[ 1\,-\,\frac{e^{-\mu r_{max}}}{4}\,+\,\frac{E}{4\mu K r_{max}}e^{-\mu r_{max}}\,-\,\frac{E^2}{\mu^2 K^2 r_{max}} \Bigg] \nonumber \\
&+&\frac{e^{-\mu b}}{E^2} \Big(1\,-\,\frac{e^{-\mu b}}{4}\Big)\Bigg\} \; ,
\eaa
where
\baa
\label{betas-12-lee-wick}
A_1 (E)&=& \frac{\mu K}{6E^2}\,\Bigg[\,\Big(1\,-\,\frac{Er_{max}}{K}\Big)^3\,-\,\Big(1\,-\,\frac{Eb}{K}\Big)^3 \Bigg] \; ,
\\
A_2 (E)&=&  -\,e^{-\mu r_{max}} \Bigg[ \frac 1E \Big( 1\,-\,\frac{e^{-\mu r_{max}}}{4} \Big)\,-\,\frac 1K \Big( r_{max}\,+\, \frac1\mu \Big) \Bigg] \; .
\eaa
\begin{figure}[t]
\centering
\includegraphics[scale=0.65]{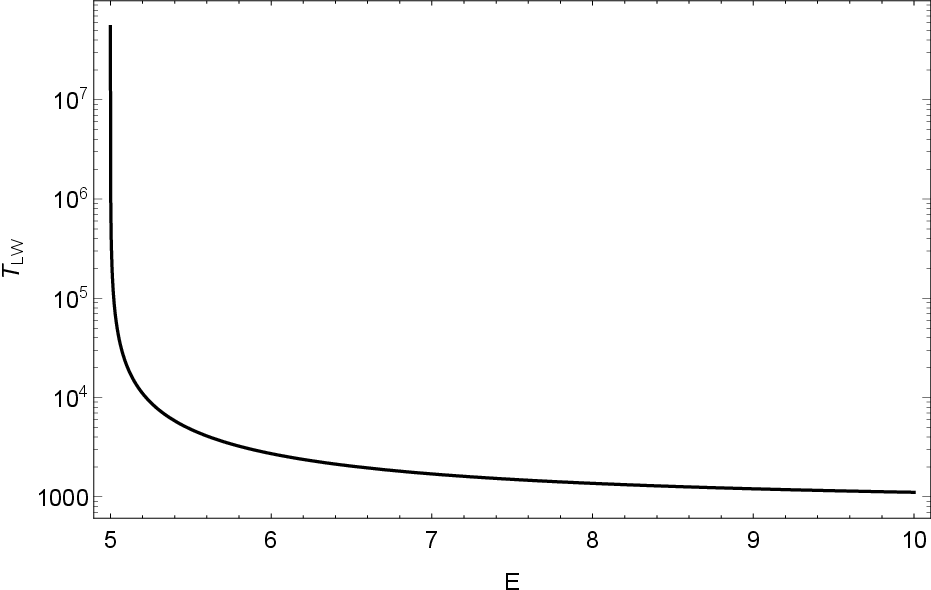}
\caption{The temperature as function of the energy for the LW potential.}\label{Ueffplasma}
\end{figure}
%

%

\subsection{The density of states and temperature for the Yukawa potential}

The Yukawa potential is obtained from subtraction of Coulomb with the LW potential in (\ref{LW}). The expression for the Yukawa potential is given by
\bee
\label{t-7}
V_{Y}(r) = K \, \frac{e^{-\mu r}}{r} \, \, ,
\eee
\noindent and using the equation \eqref{t-3}, the density of states as function of the energy is given by

\baa
\label{t-8}
g_{Y}(E) &=& \frac{AR^3 E^2}{3} \Bigg\{ r_{max}^3 \,-\, b^3\,-\, \frac{3K^2}{2E^2} \Big( e^{-2\mu r_{max}}\,-\,e^{-2\mu b} \Big) \nonumber \\
&+& \frac{6K}{E\mu} \Bigg[ e^{-\alpha r_{max}} \Big(r_{max}\,+\,\frac 1\mu \Big)\,-\,e^{-\mu b} \Big(b\,+\,\frac 1\mu \Big) \Bigg] \Bigg\} \; .
\eaa
\noindent Using the function \eqref{t-7} in the relation $V_{Y}(r=r_{max}) = E$, we obtain the maximum radial distance :
\baa
\label{t-9}
r_{max} &\approx& \frac{K}{E\,+\,\mu K}\,\,, \qquad \mbox{for} \qquad \frac{e^{-\mu b}}{b} < \frac{E}{K} < \frac{e^{-\mu R}}{R} \nonumber \\
&\mbox{and}& \qquad r_{max}\,=\,R\,\,, \qquad \mbox{for} \qquad \frac{e^{-\mu R}}{R} < \frac{E}{K} < \infty\,\,.
\eaa
Thereby, the inverse of the temperature as function of the energy is given by
\baa
\label{beta-yukawa}
\frac{1}{T_{Y}(E)} &=& \frac{2}{E^2}\,-\,\frac{AR^3 E}{K g_Y (E)}\Bigg\{ B(E)\,+\,Er^4_{max}\Big(1\,-\,\frac{6e^{-\mu R_{max}}}{Er_{max}}\Big)
\nonumber \\
&-&\frac{K^3}{E^2}\,\Bigg[ \Big( 1\,-\,\frac{Er^2_{max}}{K} \Big)\,e^{-2\mu r_{max}}\,-\,e^{-2\mu b} \Bigg] \Bigg\} \; ,
\eaa
\noindent where the function $B(E)$ is conveniently defined by
\bee
\label{betas-yukawa}
B(E)=\frac{6K}{E\mu}\,\Bigg[ e^{-\mu r_{max}} \Big( r_{max} \,+\,\frac 1\mu \Big)\,-\,e^{-\mu b} \Big( b\,+\, \frac 1\mu \Big) \Bigg] \; .
\eee
\begin{figure}[t]
\centering
\includegraphics[scale=0.70]{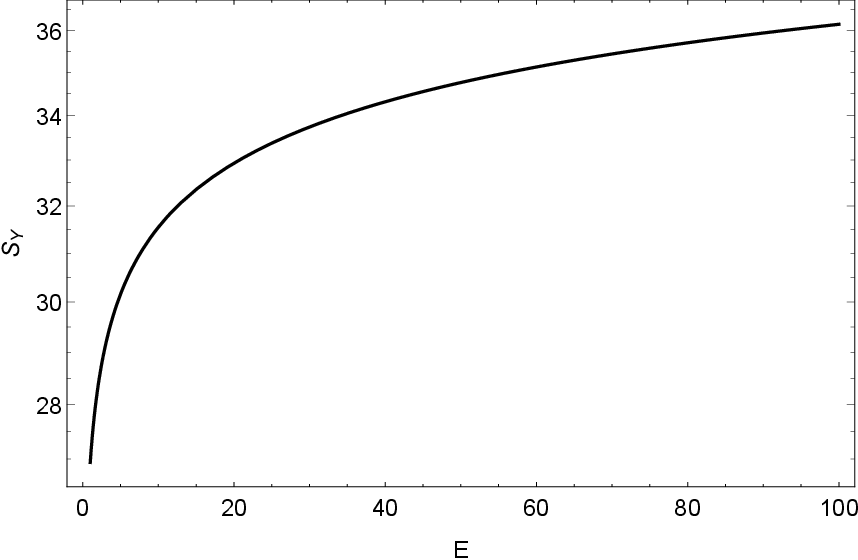}
\caption{The entropy as function of the energy for the Yukawa potential.}\label{CLW}
\end{figure}
%

%

\subsection{The density of states and temperature for Debye-H\"uckel potential}
The Debye-H\"uckel potential, also known as generalized inverse quadratic Yukawa potential, is a superposition of the Yukawa potential and the inverse quadratic Yukawa potential. This potential is asymptotic to a finite value when $r \rightarrow \infty$ and it is infinite as $r \rightarrow 0$.
%
%
%
Using the DH potential (\ref{plasmaU}) in the integral (\ref{t-7}), the density of states is given by
\baa
\label{t-11}
\frac{g_{DH}(E)}{AR^3} &=&\Bigg[\frac 13 \Big(E\,+\,V_o \Big)^3\,+\,4 V_o^2 a^3 \alpha^3 \Bigg(\frac 32 \,+\, \frac 89 V_o^2 a \Bigg) \Bigg] \Big(r_{max}^3 \,-\,b^3 \Big)
\nonumber \\
&+&\frac{4aV_o}{\alpha^2}\Big(E\,+\,V_o \Big)\Bigg[1\,+\, \alpha \Big(r_{max} e^{-\alpha r_{max}}\,-\,b\,e^{-\alpha b} \Big)\Bigg] \nonumber \\
&+&\frac{V_o a^2}{\alpha} \Big(E\,+\,3V_o \Big) \Big( e^{-2\alpha r_{max}}\,-\,e^{-2\alpha b} \Big)\,-\,a^3 V_o^2\Big(4\,+\,aV_o\Big) \ln \left(\frac{r_{max}}{b}\right) \nonumber \\
&+&4\alpha a^3 V_o^2 \Big(3\,+\,aV_o^2 \Big) \Big(r_{max}\,-\,b\Big)
\nonumber \\
\,&-&\,4\alpha^2 a^3 V_o^2 \Bigg(\frac 94 \,+\,aV_o^2 \Bigg)\,\Big(r_{max}^2\,-\,b^2\Big)\,+\, {\cal O}(r_{max}^n) \,\, ,
\eaa
\noindent where $n > 3$, and
\baa
\label{t-9}
r_{max} &\approx& \frac{a}{1\,+\,\alpha\,-\,\sqrt{\frac{E}{V_o}}}\,\,, \hspace{0.3cm} \mbox{for} \hspace{0.3cm}
-\,\Big(1\,-\,\frac{a}{b}e^{-\alpha b}\Big)^2 < \frac{E}{V_{o}} < -\,\Big(1\,-\,\frac{a}{R}e^{-\alpha R}\Big)^2 \nonumber \\
&\mbox{and}& \hspace{0.3cm} r_{max}\,=\,R\,\,, \hspace{0.3cm} \mbox{for} \hspace{0.3cm} -\,V_o\,\Big(1\,-\,\frac{a}{R}e^{-\alpha R}\Big)^2 < E < \infty\,\,.
\eaa
The inverse of the temperature is
\baa
\label{beta-plasma}
&&\frac{1}{T_{DH}(E)}\,=\, \frac{AR^3}{g_{DH}(E)}\Big\{\Big(E\,+\,V_o\Big)^2 \Big(r^3_{max}\,-\,b^3 \Big)\,+\,\frac 32 \frac{r^4_{max}}{a\sqrt{E V_o}}\,C_1(E) \nonumber \\
\,&+&\,\frac{4aV_o}{\alpha^2}\,\Big[C_2(E)\,+\, \frac{a\alpha}{4}\,C_3(E)\Big]
+\frac{2V_o (E\,+\,V_o )}{\alpha \sqrt{EV_o}} \Big(1\,-\,\alpha r_{max} \Big)\,r^2_{max}\,e^{-\alpha r_{max}} \nonumber \\
&-&\,\frac{aV_o}{\sqrt{EV_o}}\,\Big(E\,+\,3V_o \Big)\,r^2_{max}\,e^{-2\alpha r_{max}}-\frac{a^2 V_o^2 r_{max}}{2\sqrt{EV_o}} \times
\nonumber \\
\,&\times&\,\Bigg[4\,+\,aV_o\,+\,8\alpha^2 r^2_{max} \Big(\frac 94 \,+\,aV_o^2 \Big)
\,-\,4\alpha r_{max} \Big(3\,+\,aV_o^2 \Big) \Bigg] \,+\, {\cal O}\,(r^n_{max})
\Big\} \; ,
\hspace{0.5cm}
\eaa
where
\baa
\label{betas-plasma}
C_1(E) &=& \frac 13 \Big(E\,+\,V_o \Big)^3\,+\,4 V_o^2 a^3 \alpha^3\,\Big(\frac 32 \,+\, \frac 89 aV_o^2 \Big) \; ,
\nonumber \\
C_2(E) &=& 1\,+\, \alpha\,\Big( r_{max} e^{-\alpha r_{max}}\,-\,b\,e^{-\alpha b} \Big) \; ,
\nonumber \\
C_3(E) &=& e^{-2\alpha r_{max}}\,-\,e^{-2\alpha b} \; .
\eaa

\begin{figure}[t]
\centering
\includegraphics[scale=0.70]{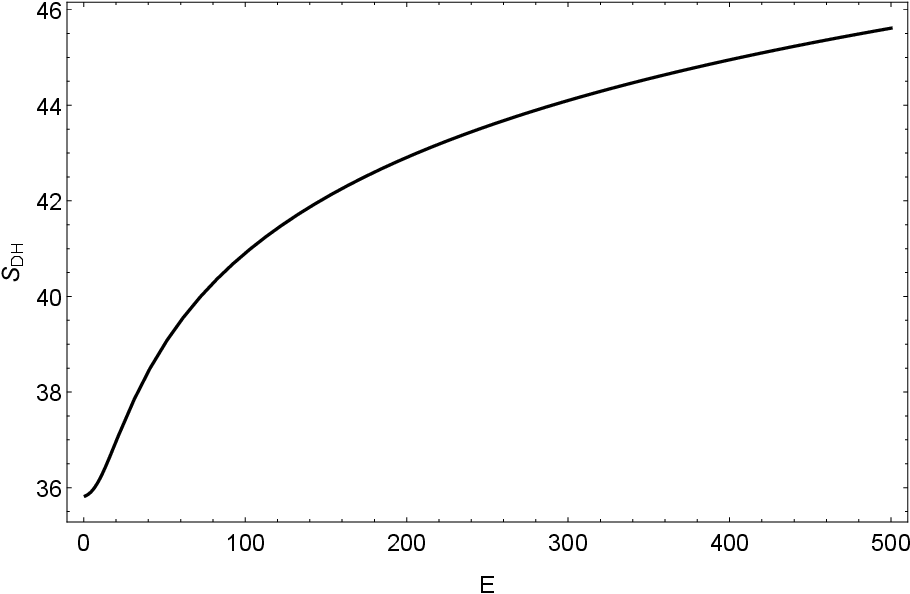}
\caption{The entropy as function of the energy for the DH potential.}\label{CLW}
\end{figure}
%


%
\section{Boltzmann-Gibbs statistical mechanics}
It is well known that systems depicted by canonical distribution cannot show negative heat capacity.   Hence, canonical distribution will head us to a completely different physical scenario for this range of energies, the mean energies.   So, it would be interesting to investigate our systems from the point of view of canonical distribution by calculating the partition function .   Therefore,
in this section, we will analyze the statistical behavior associated with the LW and plasma potentials. The statistic framework of
the models begin with the partition function. In general case, the partition function is represented by the expression
\bee\label{Zn}
Z_n = \int_{\cal R} e^{-\beta H_n} \, d^{\,n} x \, d^{\,n} p \;\; ,
\eee
where $n$ is the number of dimensions in the space, $\beta=1/T$ is the inverse of the temperature\footnote{We are assuming the unity in which the Boltzmann constant is equal to one $(k_{B}=1)$.}, $H$ is the Hamiltonian of the system, and ${\cal R}$ is the region of phase space $(x,p)$. The partition function is the term of underlying importance in the canonical ensemble.
Thereby, if we know the Hamiltonian and the phase space of the physical system, we can integrate to obtain the partition as function of temperature (T).
The mean energy of the system is obtained by the formula
\bee\label{umedio}
\langle {\cal U} \rangle = \frac{1}{Z_n} \int_{\cal R} e^{-\beta H_n} \, H_n \, d^{\,n} x \, d^{\,n} p \; ,
\eee
and the thermal capacity is defined as the derivative in relation to temperature
\begin{eqnarray}\label{CV}
C_V (T) = \frac{\partial \langle {\cal U} \rangle}{\partial T} \; ,
\end{eqnarray}
which is positive throughout range. The function $\langle {\cal U} \rangle (T)$ establishes the canonical caloric curve. The canonical ensemble can not lead to negative heat capacities. In the last section we saw that the micro-canonical scenario predicts negative heat capacities (or specific heats) together with a slow variation of energy with temperature. On the other hand, the canonical formalism predicts a phase transition with an acute variation of energy with temperature. So, it is easy to see that there is a disagreement between both formalisms \cite{padbanabahn4}.


\subsection{ The thermal capacity for Lee-Wick potential}
%

\begin{figure}
\centering
\includegraphics[scale=0.41]{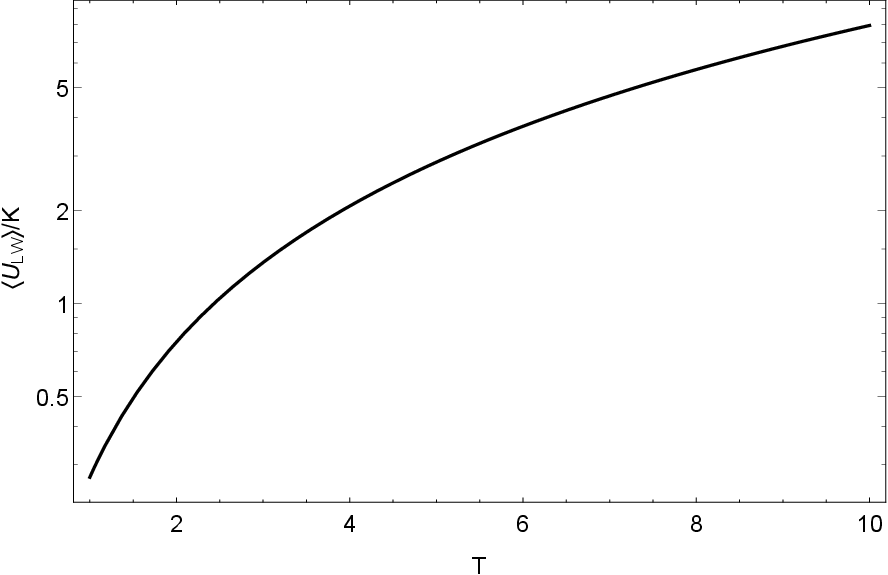}
\includegraphics[scale=0.38]{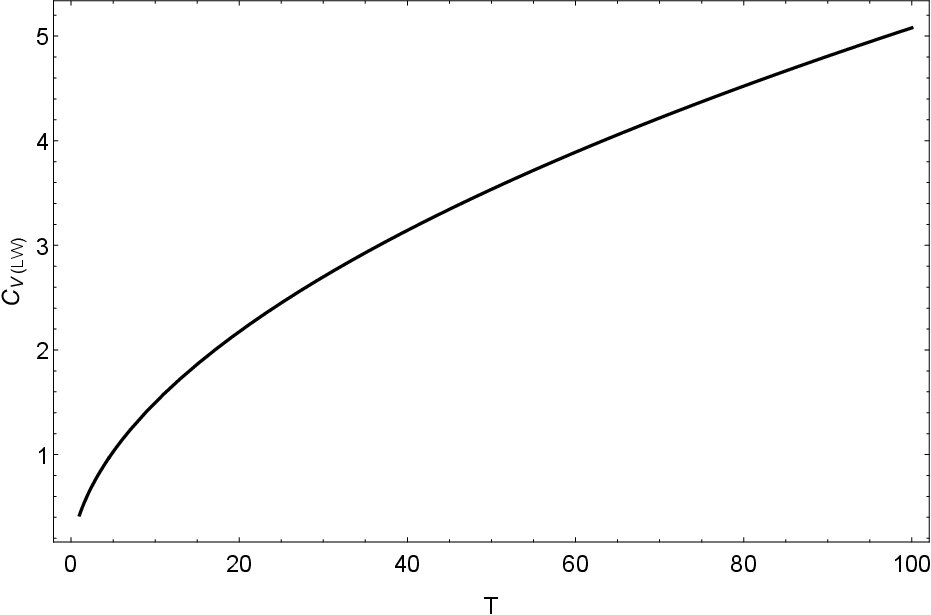}
\caption{The left panel : The mean energy by unit of $K$ associated with the LW potential as function of temperature $(T)$.
The right panel : The thermal capacity as function of the temperature for the LW potential.}\label{meanULW}
\end{figure}

%
%
%
We substitute the Hamiltonian of the two particle interacting through the LW potential into Eq. \eqref{Zn} and thus, we obtain the result
\bee\label{ZLW}
Z_{LW}(\beta)=\frac{\pi}{8} \, \sqrt{\frac{2\mu}{\beta}} \, e^{-2\sqrt{K\beta\mu}} \, \Big(3\,+2\sqrt{K \beta\mu} \Big) \; ,
\eee
where we can note that the term $K\beta\mu$ is dimensionless. Similarly, the mean energy of LW interaction as function of the temperature
is given by
\bee\label{UmeanLW}
\langle {\cal U}_{LW} \rangle(T) = KT \, \sqrt{\mu T} \, \frac{1+\sqrt{K\mu/T} }{ 3+2\sqrt{K\mu/T} } \; .
\eee
in which it is plotted in the figure (\ref{meanULW}). Using the formula (\ref{CV}) in (\ref{UmeanLW}), we obtain the thermal capacity for LW case
%
%
%
\bee
\label{thermal-capacity-lee-wick}
C_V^{LW} = \frac{3 K \sqrt{\mu T}}{2(9\,-\,4\mu K T)}\,\Big( \, 3\,-\,2\mu K T\,+\,\sqrt{\mu K T} \Big)\,+\,\frac{\mu K \sqrt{K} \, T}{2(3\,+\,2\sqrt{\mu K T} \, )^2}
\,\,.
\eee
%
%


\subsection{The thermal capacity for Yukawa potential}
We start this section with the result of partition function of a physical system of two particles interacting through a
Yukawa potential (\ref{t-7}).
%
%
Substituting Eq. \eqref{t-7} into Eq. \eqref{Zn}, we have that

\begin{figure}
\centering
\includegraphics[scale=0.32]{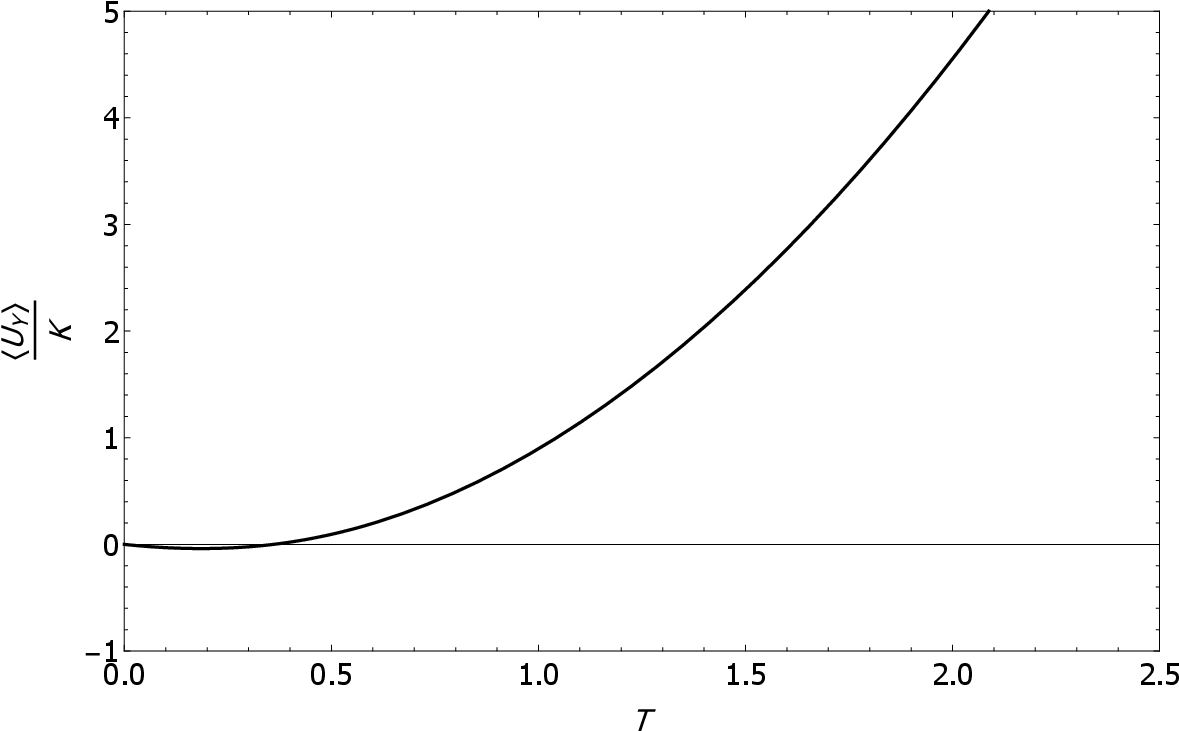}
\includegraphics[scale=0.36]{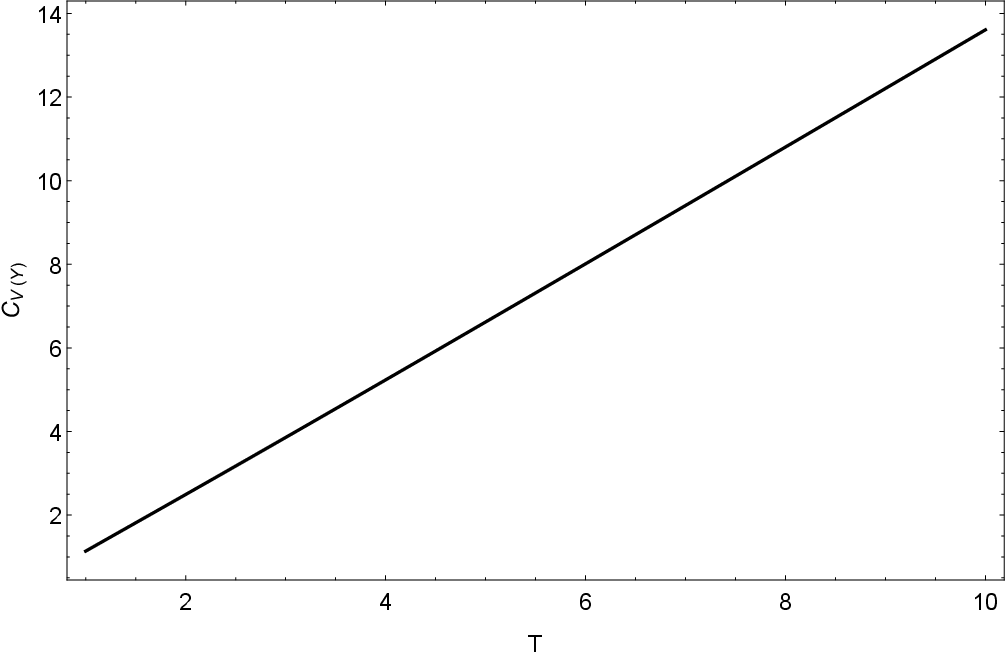}
\caption{The left panel : The mean energy associated with the Yukawa potential as function of temperature $(T)$.
The right panel : The thermal capacity as function of the temperature for the Yukawa potential.}\label{meanUY}
\end{figure}

\bee\label{ZY}
Z_{Y}(\beta) = \frac{2}{\mu} \, \sqrt{\frac{m\pi}{\beta}} \, K_{1}(\sqrt{2} K\beta\mu) \; .
\eee
The mean energy of this potential can be written as
\bee\label{UY}
\langle {\cal U}_{Y} \rangle(T) = \frac{K\mu T}{2} \, \frac{U(T)}{K_1\!\! \left(\sqrt{2} \frac{K \mu}{T} \right)} \; ,
\eee
\noindent where the function $U(T)$ is defined by
\bee \label{Ubeta}
U(T) = 2 K_{0} \left(2\sqrt{2} \, \frac{K\mu}{T}\right)-2\sqrt{2} \, K_1 \left(\sqrt{2} \, \frac{K\mu}{T}\right)
+ 2 \, K_2 \left( \sqrt{2} \, \frac{K\mu}{T} \right) \; ,
\eee
\noindent and $K_{0}$, $K_{1}$ and $K_{2}$ are Bessel functions of second kind. The mean energy in Eq. (\ref{UY}) is illustrated in Fig. (\ref{meanUY}).
From equation \eqref{CV} we have that
\bee
\label{thermal-capacity-yukawa}
C_V^{Y} = \frac{\mu K}{2}\,\frac{U(T)}{K_1(t)}\Bigg\{1\,+\,\frac{T}{K_1(t)}\Big[U'(T)\,K_1(t)\,+\,\frac{t}{2T}\,U(T)\Big(K_o(t)\,+\,K_2(t)\Big) \Big] \Bigg\}
\,\, ,
\eee
\noindent where $t=\sqrt{2}\mu K/T$ and
\bee
\label{U'}
U'(T) = -\frac{4t}{T}\Big\{K_1(2t)\,-\,\frac{\sqrt{2}}{4}\Big[K_{0}(t)\,+\,K_2(t)\Big]\,+\,\frac 12 \Big[K_1(t)\,+\,K_3(t) \Big] \Big\} \; .
\eee
%

%

%

\subsection{The thermal capacity for Debye-H\"uckel potential}
Now, we return to the potential (\ref{plasmaU}). In this case, the partition function is
\bee\label{ZP}
Z_{DH}(\beta) = \sqrt{\frac{m\pi}{2\beta}}\frac{e^{\beta(C+\alpha B)}}{B+2\alpha A}\Bigg[ \, \frac{4\sqrt{\pi}}{3}\beta\Big(B+2\alpha A\Big)^2\,+\,A \, \Bigg] \; ,
\eee
where the constants $A$, $B$ and $C$ were defined previously in Eq. (\ref{plasmaU}). The mean energy is given by
\bee\label{meanUP}
\langle {\cal U}_{DH} \rangle (\beta) = \frac{e^{C\beta}\sqrt{2m\pi\beta}}{4\,Z(\beta)\,\beta^2}\Big(\,W_1(\beta)\,+\,W_2(\beta)\,+\,W_3(\beta) \,\Big) \; ,
\eee
where the functions $U_{i} \, (i=1,2,3)$ are
\baa\label{U123}
W_1(\beta) &=& \frac{A \, e^{\alpha \beta}}{T_1^3(\beta)}\,\Big(2\alpha T_1^3(\beta)\,-\,T_1^2(\beta)\,+\,4\alpha\beta AT_1(\beta)\,+\,12\alpha\beta A \Big) \; ,
\nonumber \\
\mbox{} \nonumber \\
W_2(\beta) &=& \frac{B \, e^{\alpha\beta B}}{T_2^2(\beta)}\Big( 2\alpha \sqrt{\pi} T_2^3(\beta)\,-\,T_2^2(\beta)\,-\,\alpha\beta T_2(\beta)\,+\,2\beta A \Big) \; ,
\nonumber \\
\mbox{} \nonumber \\
W_3(\beta) &=& - \frac{C}{B} \, e^{\alpha B} \Big( T_3(\beta)\,+\,B^2 \Big) \; ,
\eaa
and
\baa\label{T123}
T_1(\beta) &=& 2\,\alpha\beta A \,+\,B \; ,
\nonumber \\
T_2(\beta) &=& \beta B(2\alpha -1) \; ,
\nonumber \\
T_3(\beta) &=& \beta A (1-2\alpha) \; .
\eaa
Using the expression in Eq. (\ref{meanUP}), we plot it as function of temperature in the Fig. (\ref{meanUP}).
\begin{figure}
\centering
\includegraphics[scale=0.55]{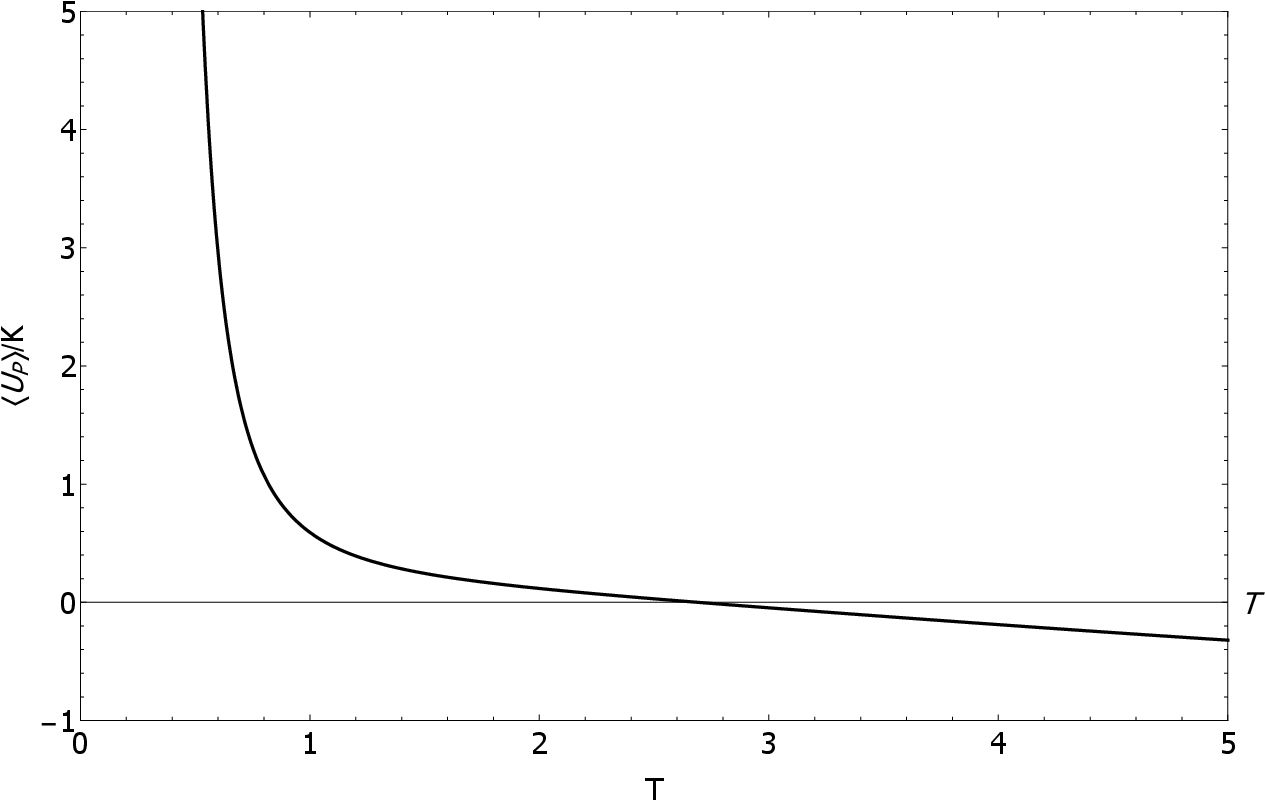}
\caption{The mean energy associated with the Debye-Hückel potential as function of temperature $(T)$.}\label{meanUP}
\end{figure}
From Eq. \eqref{CV} we have that
\bee
\label{thermal-capacity-plasma}
C_V^{DH} = -\,\frac{\sqrt{2m\pi\beta}}{4Z}\Bigg[\Big(C\,-\,\frac{3}{2\beta}\,-\,\frac{Z_1}{Z}\Big)\Big(U_1\,+\,U_2\,+\,U_3 \Big) \,+\,{U}_4\,+\,{U}_5
\,+\,{U}_6 \Bigg]
\,\, ,
\eee
where
\baa
\label{betas-plasma}
Z_1 &=&\sqrt{\frac{m\pi}{2\beta}}\,\frac{e^{\beta(C+B\alpha)}}{B+2\alpha A}\,\Bigg\{\Big[C\,+\,\alpha B\,-\,\frac{1}{2\beta}\Big]\,\Big[\frac{4\sqrt{\pi}}{3}\,\beta\Big(B\,+\,2\alpha A\Big)^2\,+\,A \Big]
\nonumber \\
&\,+&\, \frac{4\sqrt{\pi}}{3}\,\Big(B\,+\,2\alpha A\Big)^2 \Bigg\} \; ,
\nonumber \\
{U}_4 &=& \frac{A\alpha e^{\alpha \beta}}{T_1^4}\,\Big[2\alpha\,T_1^4\,-\,T_1^3\,+\,2A\,\Big(3\,+\,2\alpha\beta \Big)\,T_1^2\,
\nonumber \\
&&
+\,4A\,\Big(4\alpha\beta A\,+\,3\alpha\beta \,+\,3\Big)\,T_1\,-\,72\alpha\beta\,A^2 \Big] \; ,
\nonumber \\
{U}_5 &=& \frac{B e^{\alpha B\beta}}{T_2^2}\,\Big[ \sum_{i=1}^5 a_i\,T_2^i\,+\,2A\,-\,\alpha \beta B\Big(2\alpha\,-\,1 \Big) \Big] \; ,
\nonumber \\
{U}_6 &=& \frac{A\,C}{B}\,\Big(2\alpha\,-\,1 \Big) \,e^{\alpha B} \,\, .
\eaa
The U's and T's functions are defined in Eqs. \eqref{U123} and \eqref{T123}, respectively, and the $a$'s coefficients in ${U}_5$ are given by
\baa
\label{coeficientes-da-eq-anterior}
a_1 &=& -\,\alpha\,-\,2B(2\alpha-1)\,-\,4\alpha \beta B (2\alpha - 1) \; ,
\nonumber \\
a_2 &=& 2\,\alpha\,B\Big[\alpha\beta\,+\,3\sqrt{\pi}\,(2\alpha - 1)\,+\,\beta\,(2\alpha - 1) \; ,
\nonumber \\
a_3 &=& -\,\alpha^2\,\beta^2\,+\,2\,B\,(2\alpha-1) \; ,
\nonumber \\
a_4 &=& -\,\alpha\,B\,\Big[1\,+\,4\,(2\alpha -1)\,\sqrt{\pi} \Big] \; ,
\nonumber \\
a_5 &=& 2\,\alpha^2\,B\,\sqrt{\pi} \; .
\eaa


%


\section{Conclusions and final comments}

In spite of the two-body problem be a celestial problem, since it was designed to explain how two celestial bodies behave, the objective of this work is to analyze if, even so, it can be used as an example to small scale problems such as nuclear scale.  The same happens to Yukawa potential which was designed to describe the interaction which occurs inside the atomic nucleus.

As expected, it is not possible to use the Yukawa potential right into the two-body problem, since, as we have said, speaking in terms of the concept both have different idealizations and, in principle, incomparable ones.   However, in the application of the old thought about the subatomic particles, namely, not applying the quantum mechanical rules and assuming that it is possible to explain the motion of the nucleons beginning with the Newton laws, using the approximations and proper corrections, this became very possible.

Differently from a problem that considers both the Sun and Earth, where in general it is only the Earth that turns around the Sun, the same can be considered as true.  Therefore, when we apply the Yukawa potential, the modifications of the coordinates are necessary, since, generally speaking, the nucleons are, both of them, in motion.  In this way, the better way to describe the motion of two bodies is considering the relative position between them.   Other considerations that we have made are about the reduced mass $\mu$ and distance $r$.  The reduced mass and the distance have to be considered as something very small in order to justify the used potential.

Taking into account all these observations, the Yukawa potential describes an orbit equation analogous to the Coulomb potential, which reinforce the resemblance between both depicted at the beginning of this paper.   Nevertheless, we have considered the second order term, even we have a very small contribution, to verify the differences between the celestial orbits and the quantum orbits, when we see from a classical point of view.


\section*{Acknowledgments}

\noindent  The authors thank CNPq (Conselho Nacional de Desenvolvimento Cient\' ifico e Tecnol\'ogico), Brazilian scientific support federal agency, for partial financial support, Grants numbers 313467/2018-8 (M.J.N.) and 406894/2018-3 (E.M.C.A.). M. J. N. also thanks the Department of Physics and Astronomy at the University of Alabama for the kind and warm hospitality

\end{document}